\documentclass[twocolumn]{aastex7}
\usepackage{color}
\usepackage[titletoc]{appendix}
\usepackage{amsmath}
\usepackage{amssymb}
\usepackage{mathtools}
\usepackage{upgreek}
\usepackage{comment}
\usepackage{enumitem}
\usepackage{gensymb}
\usepackage{natbib}
\usepackage{graphicx}
\usepackage{bm}
\usepackage{totcount}
\usepackage{multirow}
\usepackage{hyperref}
\usepackage{cleveref}
\usepackage{tabularx}
\usepackage[T1]{fontenc}

\newcommand{\tc}{\tilde{c}}

\newtotcounter{citnum} 
\def\oldbibitem{} \let\oldbibitem=\bibitem
\def\bibitem{\stepcounter{citnum}\oldbibitem}

\crefname{subsection}{subsection}{subsections}
\newcommand{\ck}{\tilde{c}_k}

\shortauthors{}
\shorttitle{}

\begin{document} 

\title{Self-consistent Dynamical and Chaotic Tides in the \texttt{REBOUNDx} framework}

\author[0009-0003-3584-6698]{Donald Liveoak}
\affiliation{Department of Physics, Massachusetts Institute of Technology, Cambridge, MA 02139, USA}
\affiliation{MIT Kavli Institute for Astrophysics and Space Research, Massachusetts Institute of Technology, Cambridge, MA 02139, USA}
\affiliation{Department of Physics, University of Michigan, Ann Arbor, MI 48109, USA}
\email{dliveoak@mit.edu}

\author[0000-0003-3130-2282]{Sarah C. Millholland}
\affiliation{Department of Physics, Massachusetts Institute of Technology, Cambridge, MA 02139, USA}
\affiliation{MIT Kavli Institute for Astrophysics and Space Research, Massachusetts Institute of Technology, Cambridge, MA 02139, USA}
\email{sarah.millholland@mit.edu}

\author[0000-0002-3752-3038]{Michelle Vick}
\affiliation{Department of Physics, Brown University, Providence, RI 02912, USA}
\email{michelle_vick@brown.edu}

\author[0000-0002-9908-8705]{Daniel Tamayo}
\affiliation{Department of Physics, Harvey Mudd College, Claremont, CA 91711, USA}
\email{dtamayo@g.hmc.edu}

\defcitealias{vick2019chaotic}{V19}

\begin{abstract}
At high eccentricities, tidal forcing excites vibrational modes within orbiting bodies known as dynamical tides. In this paper, we implement the coupled evolution of these modes with the body's orbit in the \texttt{REBOUNDx} framework, an extension to the popular $N$-body integrator \texttt{REBOUND}. We provide a variety of test cases relevant to exoplanet dynamics and demonstrate overall agreement with prior studies of dynamical tides in the secular regime. Our implementation is readily applied to various high-eccentricity scenarios and allows for fast and accurate $N$-body investigations of astrophysical systems for which dynamical tides are relevant.
\end{abstract}

\section{Introduction}
\label{sec: Introduction}

When planets or stars orbit with short pericenter distances, oscillatory modes can be excited within the orbiting body in a mechanism known as dynamical tides. The energy of these oscillations comes at the expense of orbital energy. When damping is negligible, the modes are still ringing from one pericenter passage to the next, and the evolution of the modes is chaotic for high-eccentricity orbits. When the mode amplitudes become too large and dissipate due to non-linear effects, the energy exchange is irreversible, resulting in inward migration in a process known as chaotic tides. The resulting migration is much faster than in the case of equilibrium tides \citep{vick2019chaotic}.

The theory of chaotic tidal migration was first developed in the context of binary star capture \citep{mardling1995chaosb, mardling1995chaos}. \cite{ivanov2004tidal} applied this process to the tidal migration of exoplanets and developed the iterative map description of dynamical tides, which was later improved by \cite{ivanov2007dynamic}, \cite{wu2018diffusive} and \cite{vick2018dynamical}. Chaotic tidal migration has been used to explore formation channels of both low- and high-mass exoplanets \citep{vick2019tidal, teyssandier2019formation}.

The effects of dynamical tides have been studied numerically in investigations of highly eccentric binaries \citep{lai1997dynamical, vick2018dynamical}, neutron stars \citep{vick2019tidal, pratten2022impact}, and giant planet migration \citep{vick2019chaotic,teyssandier2019formation}. However, no previous study has coupled the iterative map description of dynamical tides with $N$-body integration.

In this paper, we model and implement the effect of dynamical tides in a manner consistent with $N$-body integrations through the \texttt{REBOUNDx} framework \citep{tamayo2020reboundx} as part of the \texttt{REBOUND} $N$-body package \citep{rein2012rebound}, allowing for more precise investigations of dynamical tides. Our implementation couples the fast and accurate iterative map description of dynamical tides with $N$-body orbital evolution. Although our implementation can in principle be applied to a variety of systems, this paper will focus on star-planet systems. 

The structure of the paper is as follows. In Section \ref{sec:math}, we describe the mathematical model of dynamical tides presented in \cite{vick2019chaotic}, hereafter \citetalias{vick2019chaotic}, as well as the coupling to the $N$-body equations of motion via the drag force method \citep{samsing2018implementing}. In Section \ref{sec:implementation}, we present the details of the \texttt{REBOUNDx} implementation. In Section \ref{sec:tests}, we present various tests that show overall agreement with the results of \citetalias{vick2019chaotic}. In Section \ref{sec:discussion}, we present additional applications of our implementation. We conclude in Section \ref{sec:conclusion}.

\section{Mathematical Model}
\label{sec:math}

To model the influence of dynamical tides on the evolution of the orbits of planets, we adopt the map derived in \cite{ivanov2004tidal} and refined in \citetalias{vick2019chaotic}. Specifically, we model planets as $\gamma = 2$ polytropes ($P \propto \rho^\gamma$, with $P$ pressure and $\rho$ density) and limit our discussion to the evolution of the fundamental (f-) modes, which are the most easily excited as a result of tidal forcing, given the assumed planet structure \citepalias{vick2019chaotic}.

Due to the short timescale of the periapse passage for high-eccentricity systems, dynamical tides are typically assumed to exchange energy with the orbit instantaneously at periapse \citep{ivanov2004tidal}. Over several orbits, these impulses lead to a variety of long-term behaviors, such as low-amplitude oscillations, high-amplitude resonant oscillations, and chaotic dissipation of orbital energy \citep{vick2018dynamical}.

We denote the normalized mode amplitude at the $k^\text{th}$ pericenter passage by $\ck \in \mathbb{C}$, with normalization chosen such that the mode energy is $E_k = E_{B,0}|\ck|^2$, where $E_{B,0}$ is the planet's initial orbital energy. The parameter $\Delta E$, which is the change in mode energy assuming ${\ck=0}$, is given by
\begin{equation} \label{eq:dE}
    \Delta E = \frac{G M_\star^2}{r_p^6} R_p^5 T(\eta, \sigma, e),
\end{equation}
where $G$ is the gravitational constant, $M_\star$ is the mass of the star, $R_p$ is the radius of the planet, and $r_p$ is the pericenter distance, and $e$ is the orbital eccentricity. Furthermore, $\eta = r_p/r_\text{tide}$ is the pericenter distance in units of the tidal radius $r_\text{tide} = R_p \left(M_\star/M_p\right)^{1/3}$, $M_p$ is the mass of the planet, and $\sigma$ is the frequency of the f-mode in the inertial frame. We consider the $l=m=2$ mode as it is the most important for energy transfer \citepalias{vick2019chaotic}.
The unitless function $T$ is given by
\begin{equation}
    T(\eta, \sigma, e) = 2 \pi^2\frac{\sigma}{\epsilon} Q^2{K_{22}}^2,
\end{equation}
where $\epsilon$ is a parameter with units of frequency related to the f-mode frequency, $Q$ is a dimensionless tidal overlap integral, and $K_{22}$ is an integral expression which can be approximated at high eccentricities as \citep{lai1997dynamical}
\begin{equation}
    K_{22} \approx \frac{2z^{3/2}\exp(-2z/3)}{\sqrt{15}}\left(1 - \frac{\sqrt{\pi}}{4\sqrt{z}}\right)\eta^{3/2},
\end{equation}
where $z = \sqrt{2} \sigma / \Omega_p$, with $\Omega_p$ being the rotational frequency of the planet. For the $\gamma = 2$ polytrope f-mode, we have $Q \approx 0.56$ \citepalias{vick2019chaotic}. Typically, the energy exchanged between the f-mode and the planet's orbit is greater than $\Delta E$, since when $|\tilde{c}| > 0$, the f-mode is more easily excited \citepalias{vick2019chaotic}.

For simplicity, we assume that in the high-eccentricity regimes where dynamical tides are relevant, the planet rotates at a pseudo-synchronous rate determined by
\begin{equation}
    \Omega_p = \Omega_{\text{ps}} := \frac{f_2(e)}{(1 - e^2)^{3/2} f_5(e)} n
\end{equation}
with $n$ the mean motion of the orbit and
\begin{align}
    f_2(e) &= 1 + \frac{15}{2} e^2 + \frac{45}{8} e^4 + \frac{5}{16} e^6\\
    f_5(e) &= 1 + 3e^2 + \frac{3}{8} e^4.
\end{align}
We note that while we assume a psuedo-synchronous rotation rate, we do not self-consistently evolve the spin rate of the planet (see Section \ref{sec: L conservation} for more discussion.)

\begin{figure}
    \centering\includegraphics[width=\linewidth]{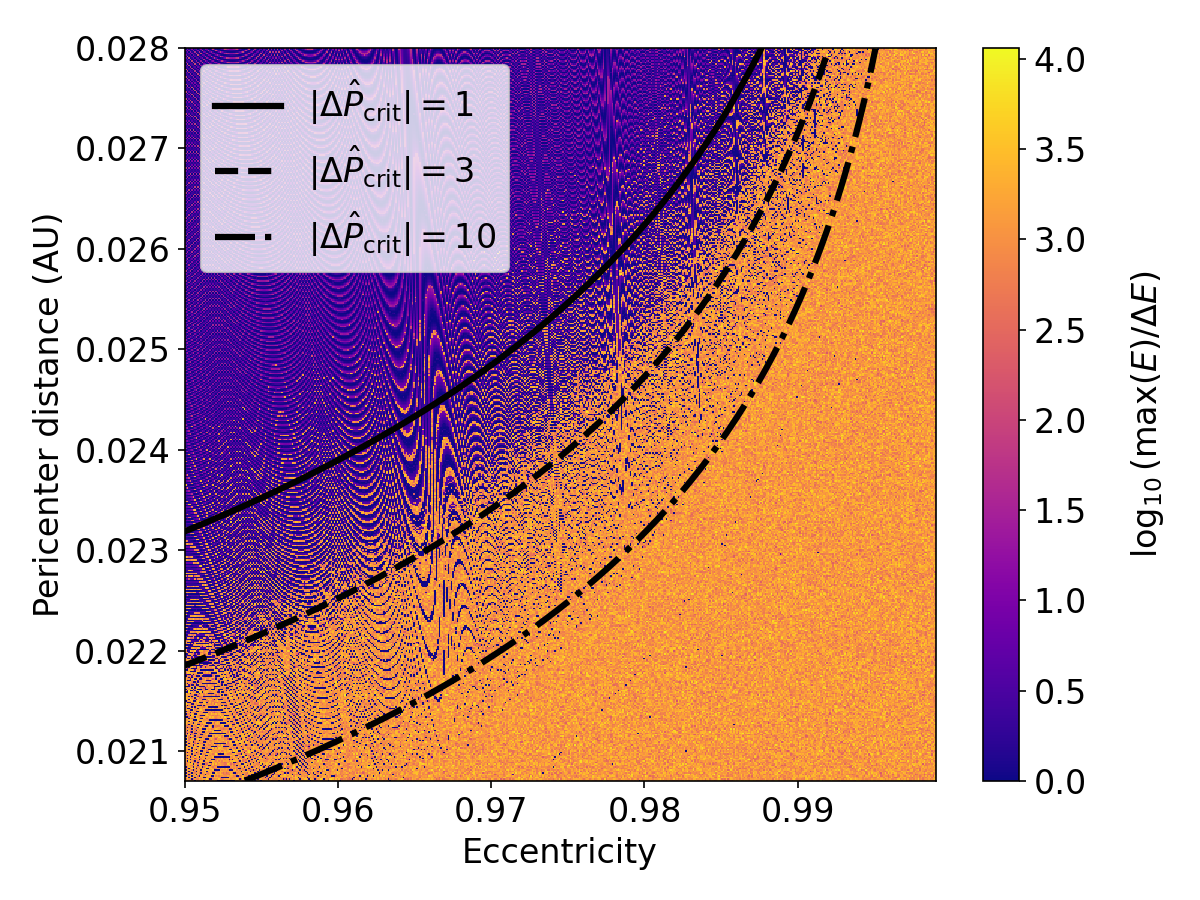}
    \caption{Mode evolution in the iterative map defined by \cref{eq:dE} evolved over $10^3$ orbits for a variety of orbital parameters. The black lines correspond to various values of $|\Delta \hat{P}_\text{crit}|$.}
    \label{fig:decoupled-map}
\end{figure}

The parameter $|\Delta \hat{P}_k|$, which can be interpreted as the change in the phase of the mode at the $k^\text{th}$ pericenter passage, is given by
\begin{equation}\label{eq:dP_hat}
|\Delta \hat{P}_k| = \frac{3}{2} \sigma_{k-1} P_{k-1} \frac{\Delta E + 2 \sqrt{E_{k-1}\Delta E}}{|E_{B, k-1}|},
\end{equation}
where $E_{B, k-1}$ is the orbital energy after the $(k-1)^{\text{th}}$ periapse passage. Equation~(\ref{eq:dP_hat}) is derived by calculating the maximum possible change in the orbital period, |$P_k - P_{k-1}$|, due to energy exchange between the orbit and planet f-mode at the $k$-th pericenter passage. This value is achieved when the f-mode is perfectly in phase with the tidal kick at pericenter. Multiplying the change in orbital period by the f-mode frequency, $\sigma_{k-1}$, yields a maximum possible phase change of the f-mode due to tidal energy transfer \citepalias{vick2019chaotic}.
For $|\Delta \hat{P}_k| \gtrsim 1$, the phase of the mode at each pericenter passage will be nearly random, and the f-mode evolution (and consequently the orbital evolution) is chaotic. On the other hand, when $|\Delta \hat{P}_k| \lesssim 1$, the mode and orbital evolution is oscillatory \citep{vick2018dynamical}. The critical value of $|\Delta \hat{P}_k|$ for chaotic behavior to occur depends on the system and can be anywhere from $\sim 0.1$ to $1$. \Cref{fig:decoupled-map} shows the outcome of the map over $10^3$ iterations for a variety of initial conditions and demonstrates agreement with \citetalias{vick2019chaotic}.

As discussed in \citetalias{vick2019chaotic}, this model often leads to large mode energies that exceed the binding energy of the planet and are physically unrealistic. Thus, we assume that if the mode energy exceeds some critical threshold $E_{\text{max}}$, it is dissipated nonlinearly over a single orbital period to some value $E_{\text{resid}}$. For a discussion of how these parameters affect the long-term f-mode evolution, see \citetalias{vick2019chaotic}.

\subsection{Dynamical tides as a drag force}
The previous analytical description of dynamical tides provides an iterative map which yields a change in mode energy per orbit of
\begin{equation}\label{eq:energy-change}
    \Delta E_k = |E_{B, 0}|(|\tc_{k-1}+\Delta \tc|^2 - |\tc_{k-1}|^2)
\end{equation}
where $\Delta \tilde{c} = \sqrt{\Delta E / |E_{B, 0}|}$ is the change in the real part of the f-mode amplitude at periapse.

We model the coupling of the f-mode evolution to the planet's orbit using the drag force model derived and implemented in \cite{samsing2018implementing}. In particular, we apply a tangential drag force $\vec{F_k} = -\mathcal{E}_k\vec{v}/r^n$, where $n \in \mathbb{Z}^+$ is an exponential parameter and $\mathcal{E}_k$ is a drag coefficient that depends on $\Delta E_k$ and the orbital elements of the system. In practice, larger values of $n$ cause the energy exchange to occur closer to periapse. We note that this method does not perfectly conserve angular momentum during the energy transfer. We discuss this point in more detail in Section \ref{sec: L conservation}.

Since $\vec{F} \parallel d\vec{s}$, we write the total change in energy over one orbit as a result of the drag force by integrating over the true anomaly $\theta$:
\begin{equation}
    \Delta E_k = -\int_{-\pi}^\pi \frac{\mathcal{E}_k v}{r^n} \frac{ds}{dt} \frac{dt}{d\theta} d \theta.
\end{equation}
Using properties of Keplerian orbits, we invert this equation to find (see \citealt{samsing2018implementing} for more details):
\begin{equation}\label{eq:drag_coef}
    \mathcal{E}_k = -\Delta E_k \cdot \frac{1}{2}\frac{[a(1-e^2)]^{n-1/2}}{[G(M_\star + M_p)]^{1/2} \mathcal{J}(e,n)},
\end{equation}
where $a$ is the semi-major axis and
\begin{equation}
    \mathcal{J}(e,n) = \int_{-\pi}^{\pi} \frac{1 + e\cos\theta - (1-e^2)/2}{(1+e\cos\theta)^{2-n}} \, d\theta.
\end{equation}
The expression for $\mathcal{J}(e,n)$ has an analytical solution for all $n \in \mathbb{Z}^+$ \citep{samsing2018implementing}. Since the force decays as $\sim1/r^n$ away from periapse, we choose a relatively large value of $n=10$ so that energy is transferred between the orbit and f-mode only when the planet is close to periapse. We have that
\begin{align}
    \mathcal{J}(e,10) = \frac{\pi}{128}&(128 + 2944e^2+10528e^4  \nonumber \\
    &+ 8960e^6 + 1715e^8 + 35e^{10}).
\end{align}

\section{Implementation in REBOUNDx}
\label{sec:implementation}

Given the mathematical model of the previous section, we now discuss our implementation in \texttt{REBOUNDx}. Our general strategy is to compute the result of the mode evolution far from periapse, where the drag force vanishes, and update the drag force parameter so that the appropriate amount of energy is exchanged between the orbit and the f-mode at the subsequent periapse passage. For simplicity, we choose to calculate all mode parameters at the apoapse passage directly preceding each periapse passage, where the magnitude of the drag force is minimized. All calculations are carried out in the units specified by the \texttt{REBOUND} simulation object. 

\begin{enumerate}
    \item At each timestep, we determine whether the planet is at apoapsis by monitoring the mean anomaly between timesteps; if $M_{\text{last}} < \pi < M$, we proceed with Step 2.
    \item We calculate $\Delta E$ and $|\Delta \hat{P}_k|$ for the subsequent pericenter passage using \cref{eq:dE} and \cref{eq:dP_hat} with the mode parameters listed in \citetalias{vick2019chaotic}. If $|\Delta \hat{P}_k| \geq |\Delta \hat{P}_\text{crit}|$, we proceed with Step 3.
    \item We calculate the energy transfer between the planet's orbit and its f-mode at periapse, given by \cref{eq:energy-change}.
    We then compute and record the drag coefficient $\mathcal{E}_k$ via \cref{eq:drag_coef}.
    \item We update the mode amplitude as 
    \begin{equation}
    \tilde{c}_k = (\tilde{c}_{k-1} + \Delta \tilde{c}) \exp(-i\sigma P_k),
    \end{equation}
    where $P_k$ is the current orbital period. If $E_{B,0}|\tilde{c}_k|^2 \geq E_{\text{max}}$, we rescale $\tilde{c}_k$ so that $E_{B,0}|\tilde{c}_k|^2 = E_{\text{resid}}$.
\end{enumerate}

At each timestep, we compute the drag force via $\vec{F_k} = -\mathcal{E}_k\vec{v}/r^n$, where $r$ and $\vec{v}$ are taken with respect to the center of mass of the planet-star system. The acceleration of the planet is updated as $\vec{a}_p \to \vec{a}_p + \vec{F_k}/M_p$ and the acceleration of the star as $\vec{a}_{\star} \to \vec{a}_{\star} - \vec{F_k}/M_\star$ so that the momentum of the center of mass is conserved.

In this implementation, the change in mode energy occurs at apoapsis but the corresponding change in orbital energy occurs at periapsis, yielding a half-orbit delay in the total simulation energy when accounting for energy stored in the f-mode. However, since the mode energy does not directly alter the trajectories of the particles in the simulation, this half-orbit delay does not have any consequences in the resulting particle dynamics.

Due to the high eccentricity of the orbits, this implementation requires an adaptive timestep integrator which is able to accurately resolve each periapse passage. Integrators with a fixed timestep, such as \texttt{WHFast} \citep{rein2015whfast}, are impractical for this effect given the very small timestep required to resolve the pericenter passage. Although all subsequent tests are completed with the \texttt{IAS15} integrator \citep{rein2015ias15}, hybrid integrators that can accurately resolve close pericenter passages, such as \texttt{TRACE}, can also be a good option \citep{lu2024trace}.

A summary of the drag force's parameters is shown in \Cref{table:params}. 

\section{Test cases}
\label{sec:tests}

In this section, we present two scenarios in which chaotic tides result in rapid migration. 

\subsection{High-eccentricity migration of an isolated planet}
The first suite of tests consists of Jupiter-sized planets initialized on orbits with high eccentricities ${e_0\in[0.97, 0.99]}$ and short pericenter distances ${r_{p,0}/\text{AU}\in[0.020, 0.025]}$. For simplicity, we neglect the effects of general relativity and equilibrium tides. Each system consists of a Jupiter-sized planet of mass ${M_p=M_J}$ and radius ${R_p = 1.6\,R_J}$ orbiting a star of mass ${M_\star=M_{\odot}}$. While this set-up is not particularly physical (since it is not clear how an isolated planet would obtain such a  high eccentricity), it is useful because it isolates the effects of dynamical tides from any other perturbations.

\begin{figure}
    \centering
    \includegraphics[width=\linewidth]{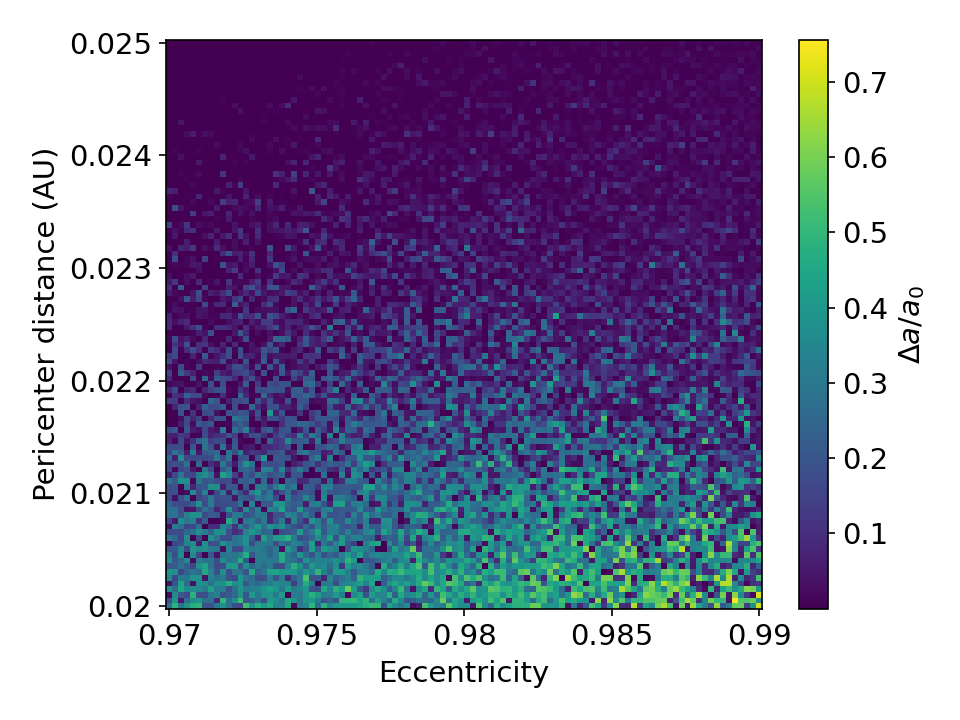}
    \caption{Chaotic migration outcomes for planets with ${M_p=M_J}$ and ${R_p=1.6\,R_J}$ orbiting a star of mass ${M_\star=M_{\odot}}$ with various initial eccentricities and pericenter distances. Bright yellow regions of the heatmap correspond to planets that migrated to a small fraction of their initial semi-major axis. Dark purple regions correspond to planets whose initial and final semi-major axes are similar.}
    \label{fig:migration-heatmap}
\end{figure}

For this region of parameter space, we sample ${N=2500}$ systems and evolve each for $10^3$ orbits. \Cref{fig:migration-heatmap} depicts the migration outcomes for each system in the sample. As expected, planets with high eccentricities and short pericenter distances tend to undergo more extreme dissipation of orbital energy and settle to narrower orbits than planets with lower eccentricities and larger pericenter distances.

\begin{figure*}[h]
    \centering
    \includegraphics[width=\linewidth]{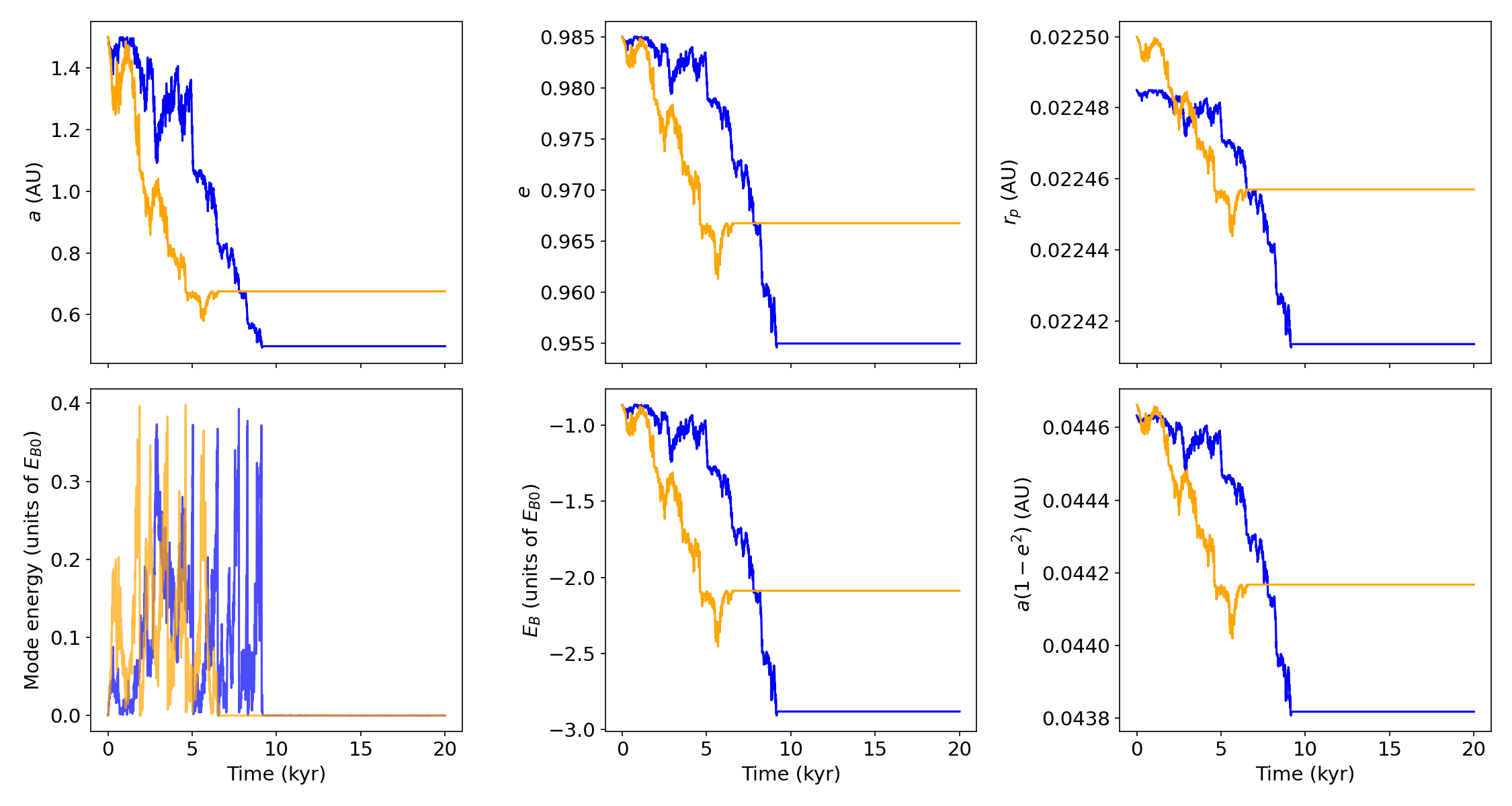}
    \caption{Two examples of high-eccentricity migration as a result of chaotic tides. Each system consists of a Jupiter-sized planet of mass ${M=M_J}$ and ${R_p=1.6\,R_J}$ orbiting a star of mass $M_{\odot}$ with initial semi-major axis ${a_0=1.5}$ AU. The orange curves correspond to initial eccentricity $e_0=0.985$ and the blue curves correspond to $e_0=0.98501$. The trajectories quickly diverge due to the chaos of the system. In this case, $E_\text{bind} = 3.8 E_{B,0}$.}
    \label{fig:high-e-migration}
\end{figure*}

\begin{figure*}[h]
    \centering
    \includegraphics[width=\linewidth]{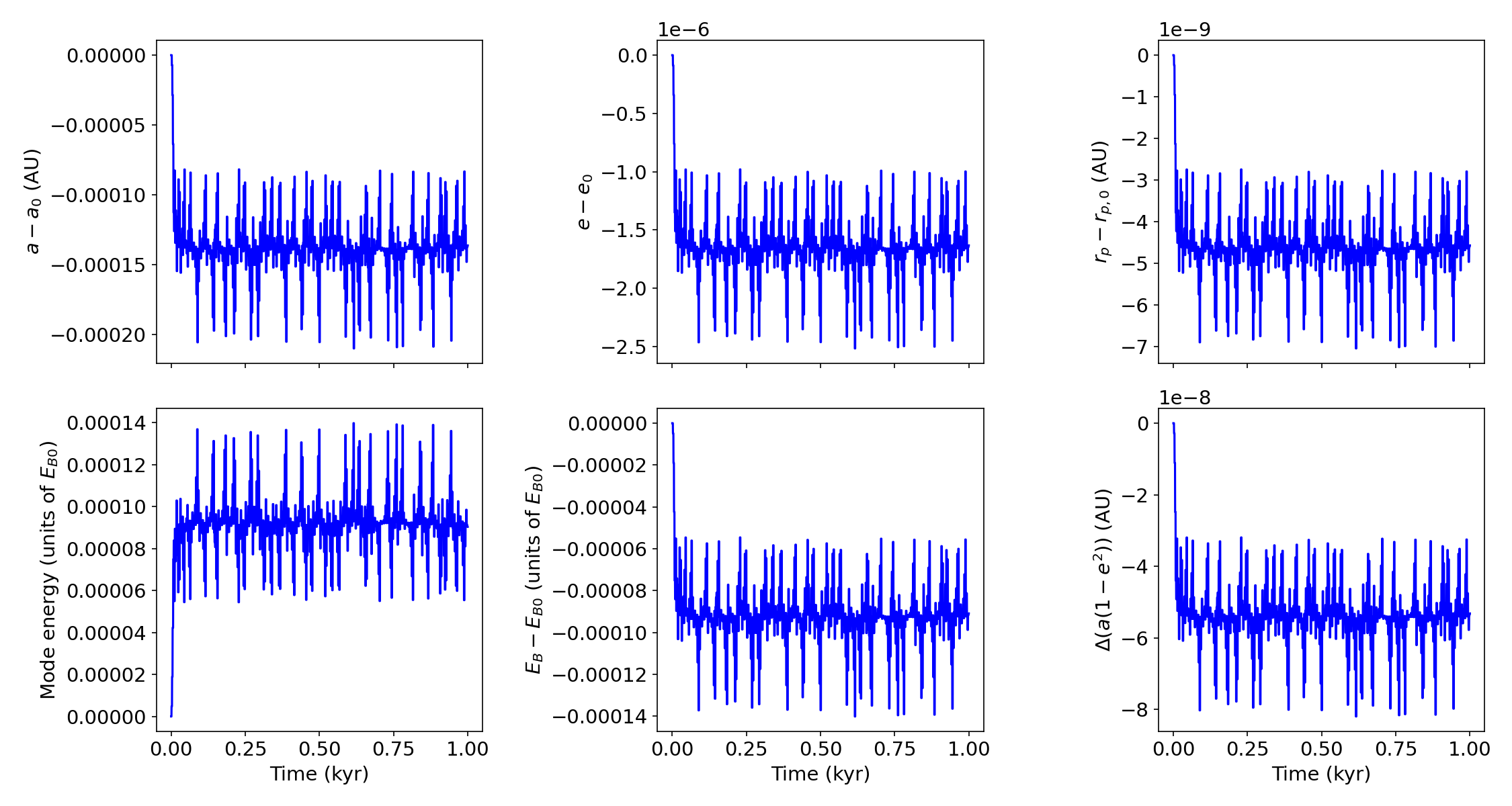}
    \caption{Example of low-amplitude oscillations as a result of dynamical tides. The system is the same as that of \cref{fig:high-e-migration}, except $e_0 = 0.982$. The trajectories are quasi-periodic.}
    \label{fig:low-amp-osc}
\end{figure*}

\Cref{fig:high-e-migration} depicts two scenarios of chaotic migration with initial semi-major axis $a_0 = 1.5$ AU and very similar initial eccentricities $e_0 = 0.985$ and $e_0 = 0.98501$. Even though the initial conditions are nearly identical, each system's evolution is vastly different, highlighting the chaotic nature of the mode evolution. A small fraction of orbital angular momentum is lost during the chaotic migration; we address this point further in Section \ref{sec: L conservation}.

\begin{figure}[h]
    \centering
    \includegraphics[width=\linewidth]{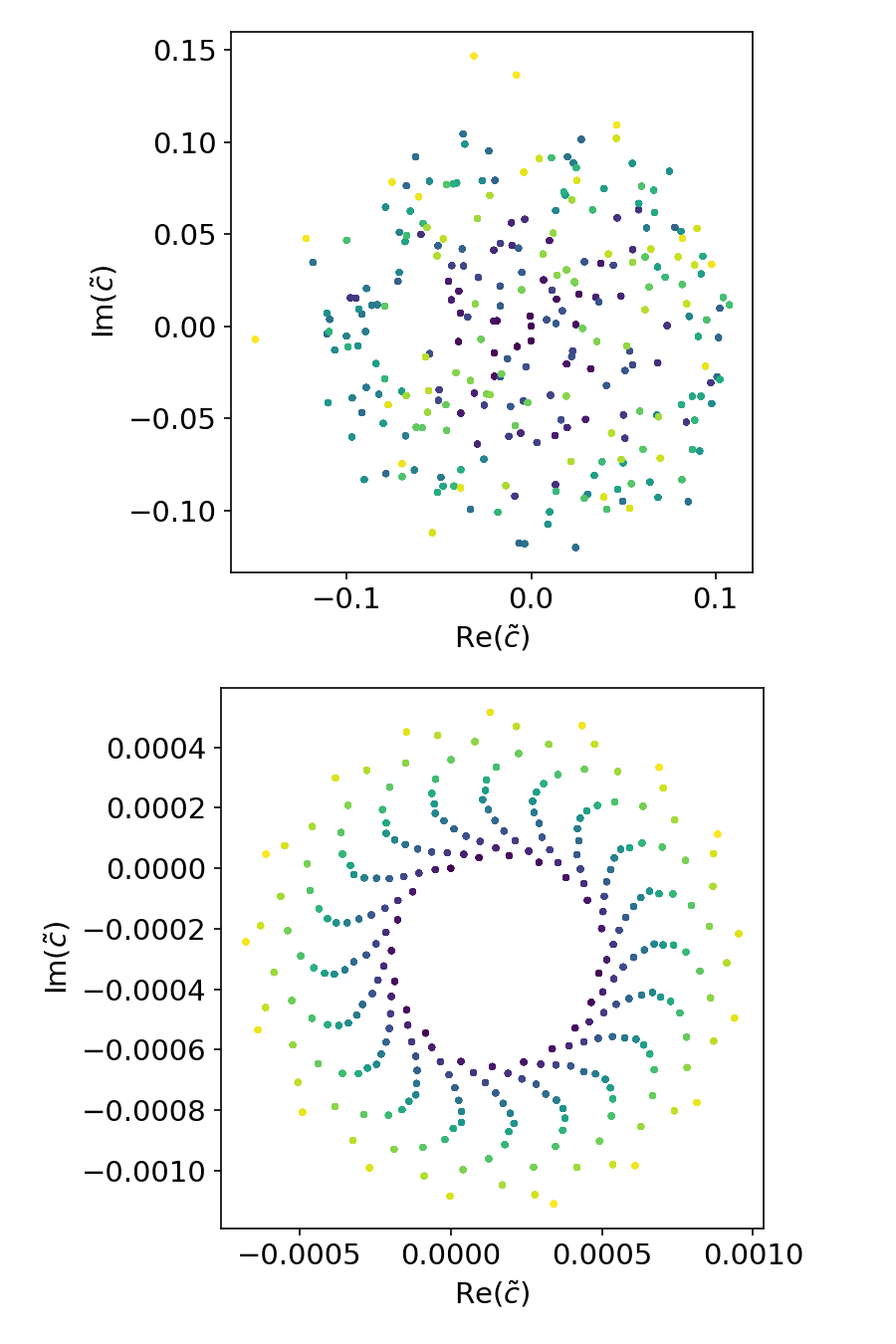}
    \caption{Evolution of f-mode over time for the same system as \cref{fig:high-e-migration} over 5 kyr, with $e_0=0.985$ (top panel) and $e_0=0.98$ (bottom panel). Points that are lighter correspond to later times in the simulation.}
    \label{fig:mode-evolution}
\end{figure}

When $a_0 = 1.5$ AU and $e_0=0.982$, a small amount of energy is initially transferred from the planet's orbit to its f-mode. Then, rather than undergoing chaotic migration, the planet's orbit and f-mode quasi-periodically exchange energy which results in low-amplitude oscillations depicted in \Cref{fig:low-amp-osc}. This behavior is due to a slightly larger pericenter distance than the previous case, yielding $|\Delta \hat{P}| \sim 10^{-2} < 1$. The difference between chaotic and non-chaotic mode evolution is also illustrated in the phase space of the f-mode (\Cref{fig:mode-evolution}), demonstrating general agreement with previous studies of dynamical and chaotic tides \citep{vick2018dynamical, vick2019chaotic}.

\subsection{High-eccentricity migration through vZLK cycles}
A more physically realistic channel for high-eccentricity migration is the von-Zeipel-Lidov-Kozai mechanism (vZLK), in which a planet is excited to a high eccentricity orbit as the result of a perturbing stellar or planetary companion with a high mutual inclination with respect to the planet. This mechanism results in periodic exchanges between the planet's eccentricity and mutual inclination  \citep{lithwick2011eccentric, naoz2016eccentric}. During the high eccentricity phases of vZLK, equilibrium tides result in dissipation of orbital energy on the Gyr timescale \citep{socrates2012super}. Eventually, when the planet migrates to a sufficiently tight orbit, equilibrium tides and general relativity dominate, quenching the effects of vZLK and causing long-term circularization. This mechanism has been studied both in the secular hierarchical approximation \citep{petrovich2015steady, liu2015suppression, anderson2016formation, naoz2016eccentric} and in $N$-body simulations \citep{lithwick2011eccentric}, and it is a dominant theory for hot Jupiter formation \citep{dawson2018origins}.

In this section, we use our $N$-body implementation to demonstrate how chaotic tidal evolution at the high-eccentricity phases of vZLK can result in rapid migration. We account for equilibrium tides and general relativity using the \texttt{tides\_constant\_time\_lag} and \texttt{gr\_potential} implementations in \texttt{REBOUNDx}, respectively \citep{tamayo2020reboundx, lu2023self}.

To compare with \citetalias{vick2019chaotic}, our test system consists of a Jupiter-sized planet of mass ${M_p=M_J}$ and radius ${R_p=1.6\,R_J}$ orbiting a star of mass ${M_\star=M_{\odot}}$ with initial semi-major axis $a_0=1.5$ AU and eccentricity $e_0 = 0.01$. The planet's tidal Love number is $k_{2p} = 0.25$, and we assume a constant time lag of $\Delta t_L = 1$ s. The system also contains a perturbing binary star of mass $M_b = M_{\odot}$ on a circular orbit with semi-major axis $a_b=200$ AU and initial mutual inclination of $i_0 = 87 \degree$ with the planet.

\begin{figure*}
    \centering
    \includegraphics[width=\linewidth]{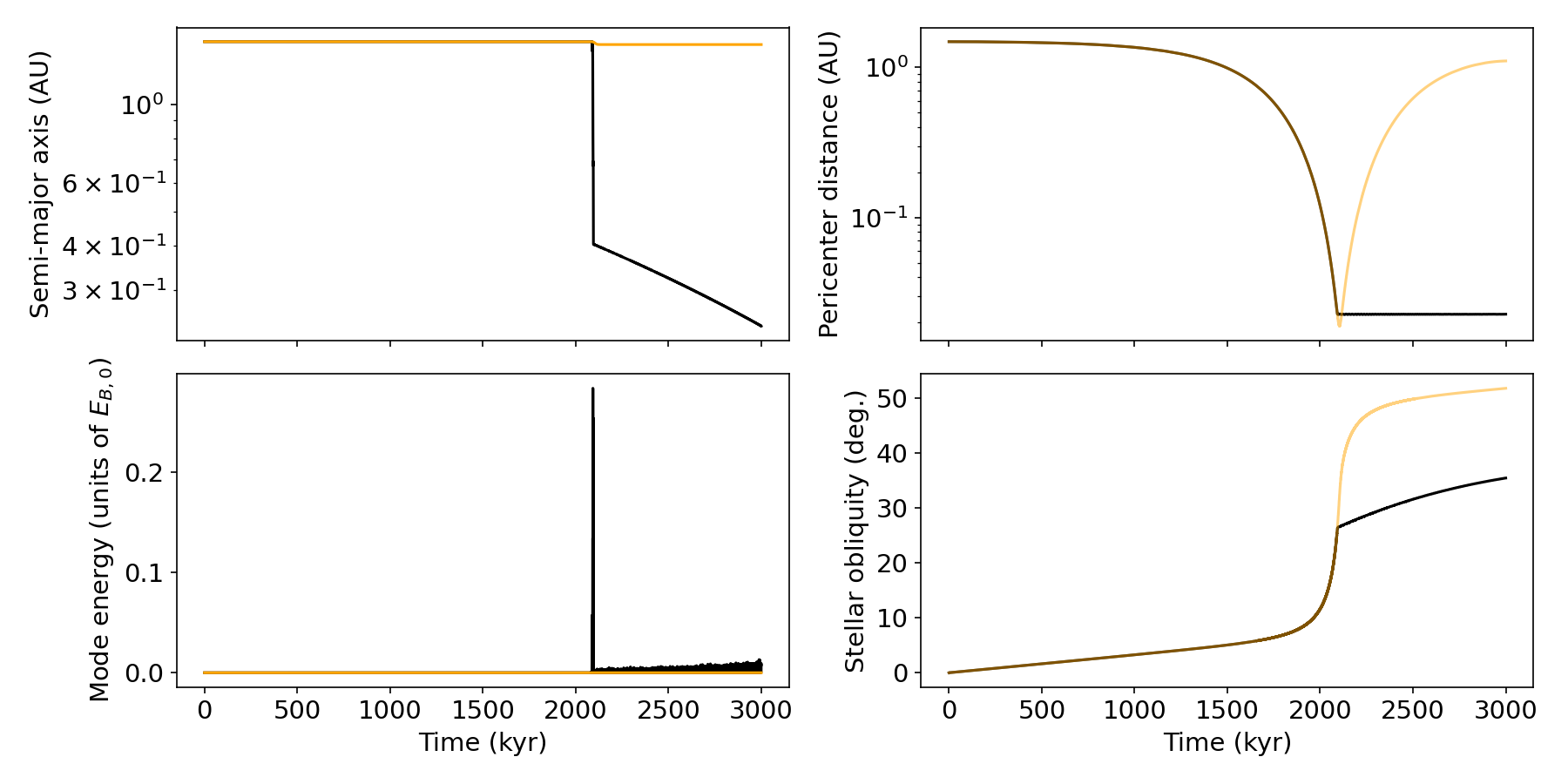}
    \caption{Time evolution of system described in Section 4.2, with dynamical tides enabled (black) and disabled (orange). In this case, $E_\text{bind} = 3.8 E_{B,0}$.}
    \label{fig:DT-ET-Kozai-inc87-ab200}
\end{figure*}

\begin{figure*}
    \centering
    \includegraphics[width=\linewidth]{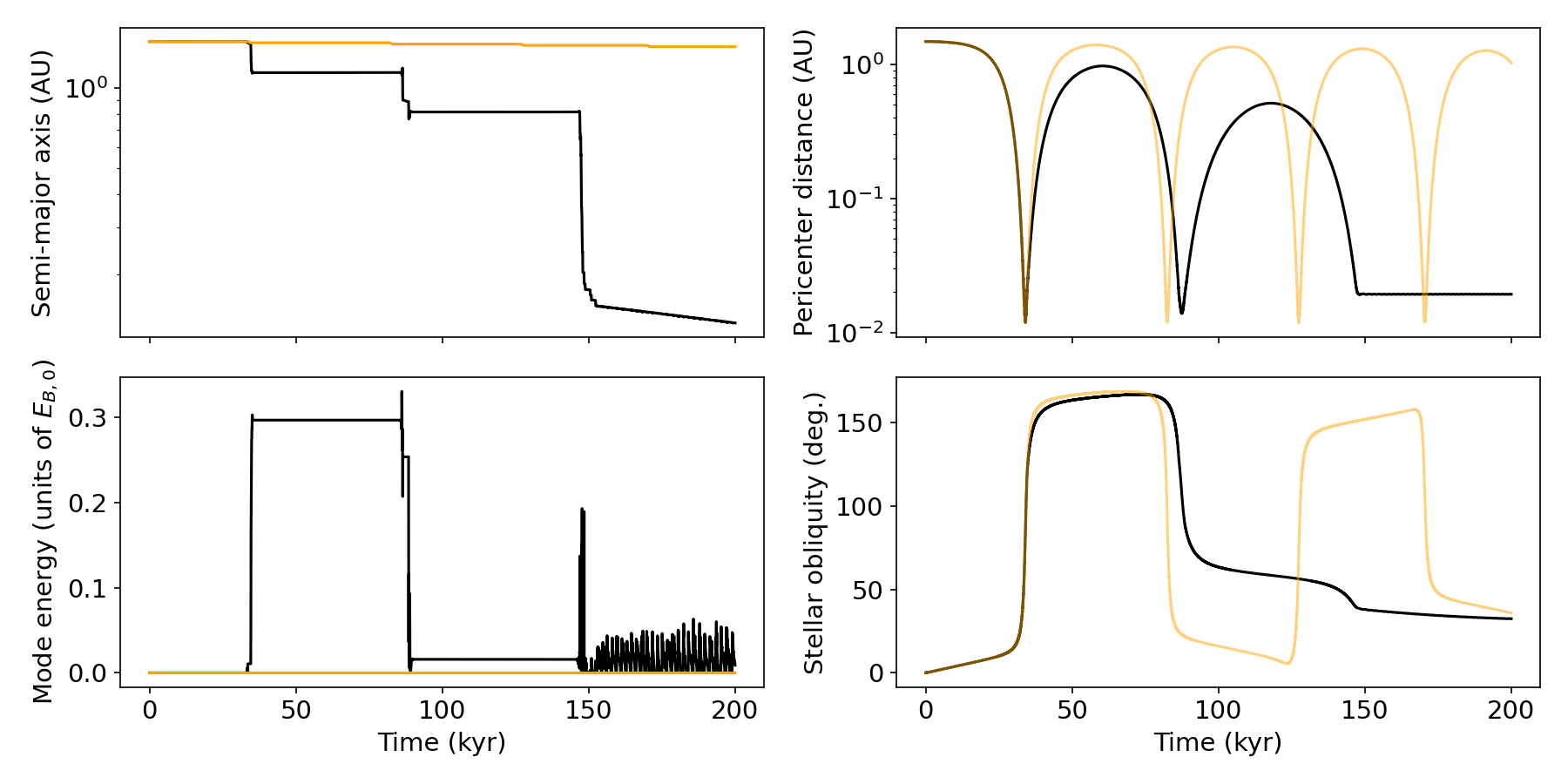}
    \caption{Time evolution of similar system as \Cref{fig:DT-ET-Kozai-inc87-ab200} with dynamical tides enabled (black) and disabled (orange), except $a_b= 50$ AU and $i_0=84.5 \degree$.}
    \label{fig:DT-ET-Kozai-inc84.5-ab50}
\end{figure*}

\Cref{fig:DT-ET-Kozai-inc87-ab200} shows the integration of these initial conditions for 3 Myr with and without the evolution of dynamical tides. In this regime, during the high-eccentricity phase of the vZLK cycles, chaotic mode evolution results in rapid migration. \Cref{fig:DT-ET-Kozai-inc84.5-ab50} shows a similar system, but with a much larger $a_0 / a_b$ and slightly smaller initial mutual inclination; in this case, chaotic mode evolution and migration occurs over a few vZLK cycles until the effects of the perturbing body are quenched by short-range forces. These two test cases demonstrate overall agreement with the findings of \citetalias{vick2019chaotic}.

\section{Discussion} \label{sec:discussion}

\subsection{Conservation of angular momentum}
\label{sec: L conservation}

As shown in \cref{fig:high-e-migration}, $\sim 1\%$ of the orbital angular momentum is lost during chaotic tidal migration, consistent with equation (13) of \cite{vick2018dynamical}. In reality, we expect a small amount of angular momentum to be transferred to both the planet's spin and vibrational modes. We note that this loss of angular momentum cannot be accounted for solely in terms of the planet spinning up, since it corresponds to an increase of $\sim0.5 \text{ hour}^{-1}$ in the planet's spin rate, which is larger than the planet's break-up frequency. To determine the consequence on the orbital elements, we compute the change in $a$ and $e$ assuming angular momentum is conserved, namely
\begin{align}
    \Delta a &= \frac{a_0(1-e_0^2)}{1-e_f^2}-a_f,\\
    \Delta e &= \sqrt{1-\frac{a_0(1-e_0^2)}{a_f}}-e_f,
\end{align}
where $a_0, a_f, e_0, e_f$ are the initial and final semi-major axis and eccentricity of the planet. We find that $|\Delta a / a_0|$ and $|\Delta e / e_0|$ are both $< 1\%$, indicating that the error in the orbital elements is negligible compared to the uncertainty introduced by the chaotic nature of the evolution. 



We emphasize that this code should be considered only in the context of tracking orbital evolution and that the spin and orbital evolution are not self-consistently coupled. We suggest that future studies derive and implement a self-consistent model for the coupled evolution of planetary spin, orbit, and mode angular momentum to enable more precise investigations of chaotic tidal migration.

\subsection{Application to other astrophysical systems}

While our code was developed and tested for high-eccentricity migration of gas giants, it can readily be adapted to diverse astrophysical setups. Dynamical tides play a critical role in the sculpting of a variety of systems with high eccentricities. By adjusting mode properties, it can be adapted to different stellar types and even to compact objects. For example, this tool could be used to study dense stellar clusters, where close encounters between stars can result in tidal capture. \citep{Fabian1975Capture, Press1977Capture, Lee1986CrossSection, Mardling2001Clusters}. After capture, dissipation due to dynamical tides may affect the rate of stellar mergers or the prevalence of systems like compact X-ray binaries \cite[e.g.][]{Ivanova2010XrayBinaries}. This tool could also be used to model the evolution of stars that are in eccentric orbits around massive black holes and will eventually undergo tidal disruption \citep{Rees1988Disruption}. 

\section{Conclusion}
\label{sec:conclusion}

In this paper, we have presented an implementation of dynamical tides as an extension to the popular $N$-body integrator \texttt{REBOUND}, through the \texttt{REBOUNDx} framework. The implementation applies the drag force method that has been used in previous investigations of tides in $N$-body integrations \citep{samsing2018implementing}. We have applied our code to the scenario of high-eccentricity migration as a formation pathway of hot Jupiters, demonstrating overall agreement with previous studies that work within secular and hierarchical approximations \citep{vick2019chaotic}.

This tool is useful for exploring astrophysical systems that exhibit extreme eccentricities and short pericenter distances in an easily accessible $N$-body framework. Though this investigation was carried out in the context of giant planet formation, our implementation could be applied in a variety of other contexts, including $N$-body studies of stellar mergers or compact X-ray binaries. Additional documentation and example Jupyter notebooks are available at {\hyperlink{https://reboundx.readthedocs.io/en/}{https://reboundx.readthedocs.io/}. \color{red}}

\section{Acknowledgments}

We thank the anonymous reviewer for their helpful comments. We also thank Tiger Lu and Mathias Michaelis for insightful discussions. This material is based upon work supported by the National Science Foundation under Grant No. 2306391. We also acknowledge support from the MIT Undergraduate Research Opportunities Program. We gratefully acknowledge access to computational resources through the MIT Engaging cluster at the Massachusetts Green High Performance Computing Center (MGHPCC) facility and the MIT SuperCloud and Lincoln Laboratory Supercomputing Center \citep{reuther2018interactive}.

\bibliographystyle{aasjournalv7}
\bibliography{main}

\newpage

\appendix
\section{Description of effect parameters}

\begin{table*}[h]
\centering
\begin{tabular}{p{0.15\linewidth}|p{0.15\linewidth}|p{0.45\linewidth}|p{0.15\linewidth}}
\hline
Parameter & Name & Description & Default Value\\
\hline
$E_\text{max}$ & \texttt{td\_E\_max}  & Maximum mode energy before non-linear dissipation & $0.1GM_p^2 / R_p$  \\
$E_\text{resid}$ & \texttt{td\_E\_resid} & Mode energy remaining after non-linear dissipation & $0.001GM_p^2 / R_p$ \\
$\text{Re}(\Tilde{c}_k)$ & \texttt{td\_c\_real} & Real component of mode amplitude & 0 \\
$\text{Im}(\Tilde{c}_k)$ & \texttt{td\_c\_imag} & Imaginary component of mode amplitude & 0 \\
$|\Delta \hat{P}_\text{crit}|$ & \texttt{td\_dP\_crit} & Critical change in phase of mode amplitude for dynamical tides to be enabled; set to 0.01 or above to only capture effects of chaotic tides & $10^{-5}$ \\
$M_{\text{last}}$ & \texttt{td\_M\_last} & Mean anomaly at last timestep; calculated internally and should not be user-specified & 0 \\
$N_\text{ap}$ & \texttt{td\_num\_apoapsis} & Number of apoapsis passages since integration has begun; primarily used for debugging & 0 \\
$t_\text{ap, last}$ & \texttt{td\_last\_apoapsis} & Simulation time of last apoapsis passage & 0 \\
$|\Delta \hat{P}_k|$ & \texttt{td\_dP\_hat} & Current change in phase of mode amplitude & 0 \\
$E_{B,0}$ & \texttt{td\_EB0} & Initial orbital energy; calculated internally and should not be user-specified & $-GM_pM_\star / (2a_0)$ \\
$\Delta E$ & \texttt{td\_dE\_last} & $\Delta E$, as computed in the most recent pericenter passage & None \\
$\mathcal{E}_k$ & \texttt{td\_drag\_coef} & Current drag coefficient as a result of dynamical tides; calculated internally and should not be user-specified & 0 \\
\end{tabular}
\caption{Planet parameters available in this implementation. The first five parameters can be specified by the user to probe different behaviors, while the latter seven are either used in the \texttt{REBOUNDx} backend or to evaluate the model and should not be modified by the user.}
\label{table:params}
\end{table*}

\end{document}